\documentclass[conference]{IEEEtran}
\IEEEoverridecommandlockouts
\usepackage{cite}
\usepackage{amsmath,amssymb,amsfonts}
\usepackage{algorithmic}
\usepackage{graphicx,subfig}
\usepackage{textcomp}
\usepackage{xcolor,color,soul,hyperref}
\usepackage{CJKutf8}
\def\BibTeX{{\rm B\kern-.05em{\sc i\kern-.025em b}\kern-.08em
    T\kern-.1667em\lower.7ex\hbox{E}\kern-.125emX}}

\begin{document}

\title{Characterizing ISCI in Multi-carrier ISAC Systems over Doubly Dispersive Channel: Joint Sensing and Communication Performance Analysis
\thanks{The work is partially supported by the National Natural Science Foundation of China (62571467, 62201162).}
}

% Impact of ISCI on ISAC Systems in Doubly Dispersive Channels: From Estimation to Capacity Analysis

% \title{On the Necessity of Handling ISCI in ISAC Systems over Doubly Dispersive Channels: A Pilot-Aided Weyl-Heisenberg Framework

\author{Xuyao Yu$^{\star}$, Zijun Gong$^{\star,*,\dagger}$, Zhilu Lai$^{\star,\dagger}$ \\
$^{\star}$The Hong Kong University of Science and Technology (Guangzhou), China\\
$^*$HKUST Fok Ying Tung Research Institute, Guangdong, China\\
$^{\dagger}$The Hong Kong University of Science and Technology, Hong Kong}
% \author{\IEEEauthorblockN{1\textsuperscript{st} Xuyao YU}
% \IEEEauthorblockA{\textit{Internet of Things Thrust} \\
% \textit{The Hong Kong University of Science and Technology (Guangzhou)}\\
% Guangzhou, China \\
% xyu991@connect.hkust-gz.edu.cn}
% \and
% \IEEEauthorblockN{2\textsuperscript{nd} Zijun Gong,\,\emph{Member}, IEEE}
% \IEEEauthorblockA{\textit{} \\
% \and
% \IEEEauthorblockN{3\textsuperscript{rd} Zhilu Lai}

\maketitle

\begin{abstract}
This paper presents a systematic analysis of inter-symbol and inter-carrier interference (ISCI) modeling in doubly dispersive channels for integrated sensing and communication (ISAC) systems. We propose a generalized OFDM (Weyl-Heisenberg) framework to evaluate four ISCI treatment approaches: (1) explicit estimation and compensation, (2) complete ignorance, (3) uncorrelated colored noise approximation, and (4) correlated colored noise modeling. Through continuous delay-Doppler channel characterization, we derive LMMSE channel estimators and corresponding estimation errors (as sensing metrics) for both pilot-assisted and fully-known symbol scenarios. The communication performance is quantified via ergodic capacity bounds under imperfect CSI. Our theoretical analysis and numerical results reveal fundamental performance-complexity trade-offs, providing insights for practical ISAC waveform and receiver design in doubly dispersive channels. 
\end{abstract}


\section{Introduction}
Integrated sensing and communication (ISAC)\cite{liuIntegratedSensingCommunications2022} systems in 6G are expected to operate in high-mobility scenarios (e.g., vehicular networks, UAVs), where the moving objects induce not only delay spread but Doppler spread as well. As a result, the channels that ISAC systems deal with are actually doubly dispersive \cite{gongChannelEstimationInterpolation2024} (also referred to as linear time-variant or doubly selective). In multi-carrier signaling based ISAC systems \cite{koivunenMulticarrierISACAdvances2024}, doubly dispersive channel may induce severe inter-symbol and inter-carrier interference (ISCI). As extensively reported in the literature \cite{tiejunwangPerformanceDegradationOFDM2006,hakobyanNovelIntercarrierInterferenceFree2018, parkIntercarrierInterferenceMitigation2024, xuHowDoesCP2025, wangCoherentCompensationBasedSensing2025}, ISCI poses a critical bottleneck that jointly impairs both sensing accuracy and communication reliability.

To characterize or mitigate ISCI’s impact, existing studies typically adopt three modeling approaches: 1) \textbf{Explicit ISCI modeling}, where advanced equalization or estimation algorithms are adopted for ISCI compensation for or exploition. For instance, \cite{keskinMIMOOFDMJointRadarCommunications2021} explicitly estimates and cancels ISCI, even leveraging inter-carrier interference (ICI) to achieve unambiguous velocity measurements. In orthogonal time frequency space modulation (OTFS) systems, ISCI is implicitly handled through delay-Doppler (DD) domain channel representation \cite{ravitejaInterferenceCancellationIterative2018,keskinIntegratedSensingCommunications2024}. 2) \textbf{Ignoring ISCI}, as in early OFDM-based ISAC systems \cite{kenneyDedicatedShortRangeCommunications2011, sturmNovelApproachOFDM2009}. This simplification is justified for low-to-moderate SNR regimes where ISCI is negligible \cite{xuHowDoesCP2025}, while enabling low-complexity designs. 3) \textbf{Modeling ISCI as Gaussian noise} \cite{parkIntercarrierInterferenceMitigation2024,wangCoherentCompensationBasedSensing2025}, which facilitates tractable performance analysis by quantifying ISCI’s impact via the signal-to-interference-plus-noise ratio (SINR).

Prior works primarily focus on task-specific ISCI analysis (e.g., OFDM radar ranging, bit-error-rate optimization) or dive into complex mitigation algorithms, but lack a unified framework to evaluate how ISCI modeling choices fundamentally limit joint sensing-communication performance. This paper addresses a pivotal trade-off: Explicit ISCI estimation maximizes accuracy but demands high computational complexity, whereas simplified ISCI models reduce complexity at the cost of performance degradation. How much performance loss is incurred by simplification? Answering this question is essential for practical ISAC system design.

This paper makes the following contributions: We propose a unified framework using generalized OFDM (also known as Weyl-Heisenberg system)\cite{kozekNonorthogonalPulseshapesMulticarrier1998} with continuous Delay-Doppler channel representation\cite{belloCharacterizationRandomlyTimevariant1963} to analyze four ISCI models (explicit, ignorant, uncorrelated/correlated colored noise), evaluating their impact through the error of linear minimum mean square error (LMMSE) channel estimation (sensing metric) and ergodic capacity under imperfect CSI (communication metric) for both pilot-only and fully-known symbol cases, providing crucial insights for practical ISAC system design.
% (as shown in Figure~\ref{fig:ISACDiagram})
% \vspace{-3mm}
% \begin{figure}[h]
%     \centering
%     \includegraphics[width=0.7\linewidth]{figure/ISACDiagram.pdf}
%     \label{fig:ISACDiagram}
%     \vspace{-5mm}
%     \caption{ISAC Diagram}
% \end{figure} 

Section II derives the signal model and formalizes the four ISCI models. Section III derives the LMMSE channel estimators for each model under both partially and fully known symbol scenarios, along with sensing error analysis. Section IV presents the ergodic capacity analysis under imperfect CSI, including upper and lower bounds for simplified models. Section V provides numerical results and comparative performance evaluation of the proposed models.

\section{Signal Model}
\subsection{General Input-output Relationship}
% The transmitted signal of the generalize-OFDM is constructed through time-frequency shifted versions of a prototype pulse function $g_t(t)$. This representation offers compatibility with modern multi-carrier waveforms including OFDM, OFDM inspired waveforms\cite{farhang-boroujenyOFDMInspiredWaveforms2016} and FBMC\cite{fettweisGFDMGeneralizedFrequency2009,zhangFBMCSystemInsight2016}. The transmitted signal is defined as:
The transmitted signal of the generalize-OFDM is constructed through time-frequency shifted versions of a prototype pulse function $g_t(t)$. This representation offers compatibility with modern multi-carrier waveforms. The transmitted signal is defined as:
\begin{equation}
s(t)=\sum_{m,n}\mathbf{X}[m,n]g_t(t-nT)e^{j2\pi mF(t-nT)},
\end{equation}
where $T$ denotes the symbol duration, $F$ represents the subcarrier spacing, and $\mathbf{X}[m,n]$ are the transmitted symbols. 

The received signal is modified by Delay-Doppler spreading function $V(\tau,\nu)$ and corrupted by an additive Gaussian noise $n(t)$
\begin{equation} 
    \begin{aligned}
    r(t) & =\iint s(t-\tau)V(\tau,\nu)e^{j2\pi\nu(t-\tau)}d\nu d\tau + n(t).
    \end{aligned}
\end{equation}

We assume that the channel process is wide-sense stationary in time and uncorrelated in delay, which is the so-called wide-sense stationary uncorrelated scattering (WSSUS) assumption. This assumption means
$$
\begin{aligned}
    \mathbb{E}[V(\tau_{1},\nu_{1})V^{*}(\tau_{2},\nu_{2})] 
    & = {\mathcal{P}}_{\mathcal{D}}(\tau_{1},\nu_{1}) \delta(\tau_{1}-\tau_{2})\delta(\nu_{1}-\nu_{2}),
\end{aligned}
$$
where ${\mathcal{P}}_{\mathcal{D}}(\tau,\nu)$ is called the \textit{scattering function} of the channel. We also assume that the scattering function has limited Doppler shift and delay, which implies that ${\mathcal{P}}_{\mathcal{D}}(\tau,\nu)$ is supported on a rectangle of spread $\Delta_{\mathcal{D}}=\tau_{\mathcal{D}}\nu_{\mathcal{D}}$, i.e.,
$$
    {\mathcal{P}}_{\mathcal{D}}(\tau,\nu)=0, \quad \text{for } (\tau,\nu) \notin \left[-\frac{\tau_{\mathcal{D}}}{2},\frac{\tau_{\mathcal{D}}}{2} \right]\times\left[-\frac{\nu_{\mathcal{D}}}{2},\frac{\nu_{\mathcal{D}}}{2}\right].
$$

The received symbols are obtained through projection onto a set of basis functions:
\begin{equation}
\mathcal{G}_{m,n}(t) = g_r(t-nT) e^{j2\pi mF(t-nT)},
\end{equation}
where $g_r(t)$ is the receive prototype pulse satisfying $\int g_t(t) g^*_r(t) dt=1$. The demodulated symbol at the $(m,n)$-th time-frequency position is:
\begin{equation}
\mathbf{Y}[m,n] = \int r(t) \mathcal{G}^*_{m,n}(t)dt.
\label{equ:InnerProduct}
\end{equation}

After reformulating \eqref{equ:InnerProduct}, we obtain the input-output relationship:
\begin{equation} 
    \begin{aligned}
        \mathbf{Y}[m,n] & = \sum_{\underline{m},\underline{n}}\mathbf{X}[\underline{m},\underline{n}]\mathbf{H}_{m,n}[\underline{m},\underline{n}]  + \mathbf{N}[m,n],
    \end{aligned}
    \label{equ:OGSignalModel}
\end{equation}
where $\mathbf{N}[m,n]\overset{\text{i.i.d.}}{\sim}\mathcal{CN}(0,\sigma^2_n)$, $\mathbf{H}_{m,n}[\underline{m},\underline{n}]$ denotes the channel coefficient from the $(\underline{m},\underline{n})$-th transmitted symbol to the $(m,n)$-th received symbol and it is expressed as
\begin{equation}
    \begin{aligned}
        & \mathbf{H}_{m,n}[\underline{m},\underline{n}] = e^{-j2\pi \underline{m} \underline{n}TF} \iint_{\mathcal{D}}  e^{j2\pi(\nu+\underline{m}F)(nT-\tau)} \\
        & V(\tau,\nu) C_{g_t,g_r}(\tau+\underline{n}T-nT,\nu+\underline{m}F-mF)d\nu d\tau, \\
    \end{aligned}
    \label{equ:ChannelCoefficientsFullModel}
\end{equation}
with $C_{g_t,g_r}(\tau,\nu)$ being the cross-ambiguity function between $g_t(t)$ and $g_r(t)$:
\begin{equation}
    C_{g_t,g_r}(\tau,\nu)=\int g_t(t)g^*_r(t+\tau)e^{j2\pi\nu (t+\tau)}dt.
\end{equation}

The coefficients $\mathbf{H}_{m,n}[\underline{m},\underline{n}]$ in \eqref{equ:ChannelCoefficientsFullModel} represent:
\begin{itemize}
\item \textbf{ISCI components} when $(m,n) \neq (\underline{m},\underline{n})$,
\item \textbf{Direct channel coefficients} when $(m,n) = (\underline{m},\underline{n})$ (representing the aligned time-frequency channel response).
\end{itemize}

\subsection{ISCI Modeling}

We present a hierarchical modeling framework for ISCI characterization, comprising four distinct approaches with varying complexity:
\begin{enumerate}
    \item \textbf{Full Model}: Maintains complete interference structure by explicitly modeling all channel coefficients.
    \item \textbf{ISCI-free Model}: Neglects all interference components, considering only direct channel coefficients.
    \item \textbf{ISCI-iN Model}: Aggregates interference and model it as independent colored noise, uncorrelated with the direct channel coefficients.
    \item \textbf{ISCI-dN Model}: Treats aggregated interference as dependent colored noise, correlated with the direct channel coefficients.
\end{enumerate}

In the remainder of this section, we present mathematical formulations for the four aforementioned models, with particular emphasis on their corresponding vectorized representations.

\subsubsection{Full Model Formulation}
The vectorized representation of the complete system derives from \eqref{equ:OGSignalModel}:
\begin{equation}
    \mathbf{y}_{fm} =\mathbf{H}_{fm}  \mathbf{x} +\mathbf{n},
\end{equation}
where the subscript $(\cdot)_{fm}$ denotes the full model, $\mathbf{y}_{fm}=\mathrm{vec}(\mathbf{Y})$, $\mathbf{x}=\mathrm{vec}(\mathbf{X})$, $\mathbf{n} = \mathrm{vec}(\mathbf{N})$. $\mathbf{H}_{fm}$ is the full channel matrix that captures all interference relationships: 
\begin{equation}
    \mathbf{H}_{fm} =
    \begin{bmatrix}
    \mathrm{vec}(\mathbf{H}_{0,0})^T \\
    \mathrm{vec}(\mathbf{H}_{1,0})^T \\
    \vdots \\
    \mathrm{vec}(\mathbf{H}_{M-1,N-1})^T
    \end{bmatrix}.
\end{equation}

Alternatively, for the convenience of channel estimation, the vectorized channel model can be rewritten as
\begin{equation}
    \mathbf{y}_{fm} = \mathbf{A} \mathbf{h} + \mathbf{n},
\end{equation}
where $\mathbf{A} = \mathbf{I}_{MN} \otimes \mathbf{x}^T$ and $\mathbf{h} = \mathrm{vec}(\mathbf{H}_{fm}^T)$. 

In this paper, we assume the transmitted symbol $\mathbf{x}$ can be decomposed as:
\begin{equation}
    \mathbf{x} = \mathbf{x}_p + \mathbf{x}_d,
\end{equation}
where $\mathbf{x}_p$ and $\mathbf{x}_d$ are zero-padded pilot and data vectors, respectively. Let $\mathcal{I}_p$ and $\mathcal{I}_d$ denote the index sets of pilot and data symbols such that $\mathcal{I}_p\cup \mathcal{I}_d = \{0, \dots, MN-1\}$ and $\mathcal{I}_p\cap\mathcal{I}_d=\emptyset$,
$$
\mathbf{x}_p[k] = \begin{cases} 
    \substack{\text{pilot}\\\text{symbol}} & k \in \mathcal{I}_p \\ 
    0 & \text{otherwise} 
\end{cases},\
\mathbf{x}_d[k] = \begin{cases} 
    \substack{\text{data}\\\text{symbol}} & k \in \mathcal{I}_d \\ 
    0 & \text{otherwise} 
\end{cases}.
$$
Moreover, the pilot symbols are assumed to be of constant modulus with amplitude $\sigma_p$ and the data symbols are assumed to be i.i.d. complex Gaussian with zero mean and $\sigma^2_d$ variance. Thus, $\mathbf{A} = \mathbf{A}_d + \mathbf{A}_p$ where $\mathbf{A}_d=\mathbf{I}_{MN} \otimes \mathbf{x}_d^T$ and $\mathbf{A}_p=\mathbf{I}_{MN} \otimes \mathbf{x}_p^T$.

To facilitate subsequent discussion, we define a data mask vector:
\begin{equation}
    \mathbf{c}_d[k] = \begin{cases} 
        1 & \text{if } k \in \mathcal{I}_d \\ 
        0 & \text{otherwise} 
    \end{cases}.
\end{equation}

\subsubsection{Simplified Models}

\paragraph{ISCI-free model} ISCI-free model assumes that when $m\neq \underline{m}$ or $n\neq \underline{n}$, $\mathbf{H}_{m,n}[m,n]=0$. Thus, by denoting $\mathbf{G}[m,n]=\mathbf{H}_{m,n}[m,n]$, the channel model can be expressed as 
\begin{equation}
    \mathbf{Y}_{fr} = \mathbf{G} \odot \mathbf{X} + \mathbf{N}.
\end{equation}

It's vectorized form can be written as
\begin{subequations}
\begin{align}
\mathbf{y}_{fr} =& \mathbf{B} \mathbf{g} + \mathbf{n}, \\
\mathbf{y}_{fr} =& \mathbf{H}_{fr}  \mathbf{x} +\mathbf{n},
\end{align}
\end{subequations}
where the subscript $(\cdot)_{fr}$ denotes the ISCI-free model, $\mathbf{B}=\mathrm{Diag}(\mathbf{x})$, $\mathbf{g}=\mathrm{vec}(\mathbf{G})$, $\mathbf{H}_{fr} = \mathrm{Diag}(\mathbf{g})$ and $\mathrm{Diag}(\mathbf{x})$ denotes a diagonal matrix whose main diagonal elements are the entries of the vector $\mathbf{x}$. $\mathbf{B}$ can also be expressed as sum of the data part $\mathbf{B}_d=\mathrm{Diag}(\mathbf{x}_d)$ and pilot part $\mathbf{B}_p=\mathrm{Diag}(\mathbf{x}_p)$. 

\paragraph{ISCI-iN model} 
% In this approach, the ISCI does not explicitly represented in the signal model by linear combination of interference channel coefficient and transmitted symbol. 
The ISCI-iN model is expressed as
\begin{subequations}
\begin{align}
    \mathbf{y}_{iN} =& \mathbf{B}\mathbf{g} + \boldsymbol{\delta}_{iN} + \mathbf{n},\\
    \mathbf{y}_{iN} =& \mathbf{H}_{iN}\mathbf{x}+ \boldsymbol{\delta}_{iN} +\mathbf{n},
\end{align}
\end{subequations}
where the subscript $(\cdot)_{iN}$ denotes the ISCI-iN model, $\boldsymbol{\delta}_{iN}$ is the colored noise term and $\mathbf{H}_{iN} = \mathrm{Diag}(\mathbf{g})$. Denote $\tilde{\mathbf{h}}=\mathrm{vec}(\mathbf{H}_{fm}^T-\mathbf{H}_{fr}^T)$, we have $\boldsymbol{\delta}_{iN}= \mathbf{A}\tilde{\mathbf{h}}$. More importantly, $\boldsymbol{\delta}_{iN}$ is assumed to be uncorrelated with the direct channel coefficients $\mathbf{g}$.

\paragraph{ISCI-dN model} Two vectorized representations of ISCI-dN model are as follows:
\begin{subequations}
\begin{align}
    \mathbf{y}_{dN} =& \mathbf{B}\mathbf{g} + \boldsymbol{\delta}_{dN} + \mathbf{n},\\
    \mathbf{y}_{dN} =& \mathbf{H}_{dN}\mathbf{x}+ \boldsymbol{\delta}_{dN} +\mathbf{n},
\end{align}
\end{subequations}
where the subscript $(\cdot)_{dN}$ denotes the ISCI-dN model, $\boldsymbol{\delta}_{dN}= \mathbf{A}\tilde{\mathbf{h}}$ and $\boldsymbol{\delta}_{dN}$ is modeled as dependent colored noise (correlated with the direct channel coefficients $\mathbf{g}$).

\section{LMMSE Sensing}
\label{sec:LMMSESensing}
This section derives the LMMSE estimates of the channel coefficients under the four proposed models, along with their corresponding estimation errors.

\subsection{LMMSE Sensing with Partially Known Transmitted Symbols}
We consider a practical scenario where the receiver has knowledge of the pilot symbols and the autocovariance of the data symbols. The LMMSE estimate of the channel coefficients for the full model is given by:
\begin{equation}
    \hat{\mathbf{h}}_{fm} = \mathbf{C}_{\mathbf{h}\mathbf{y}_{fm}} \mathbf{C}_{\mathbf{y}_{fm}}^{-1}\mathbf{y}_{fm},
    \label{equ:LMNMSEFullModelPartialKnown}
\end{equation}
where the covariance matrices are defined as:
\begin{subequations}
    \begin{align}
        \mathbf{C}_{\mathbf{h}\mathbf{y}_{fm}} =& \mathbf{C}_{\mathbf{h}} \mathbf{A}_p^H,  \\
\mathbf{C}_{\mathbf{y}_{fm}} =& \mathbf{A}_p \mathbf{C}_{\mathbf{h}} \mathbf{A}_p^H + \sigma_d^2 \mathbf{T}(\mathbf{C_{h}}) + \mathbf{C}_{\mathbf{n}}. \label{equ:Cyfm}
    \end{align}
\end{subequations}
Here, $\mathbf{C}_{\mathbf{h}}$ denotes the autocovariance matrix of the channel coefficient $\mathbf{h}$, $\mathbf{C}_{\mathbf{n}}=\sigma_n^2\mathbf{I}$ is the covariance matrix of noise $\mathbf{n}$, $\mathbf{T}: \mathbb{C}^{(MN)^2 \times (MN)^2} \to \mathbb{C}^{MN \times MN}$ is a matrix compression operator that maps a block-partitioned matrix $\mathbf{C} = [\mathbf{C}_{ij}]_{i,j=1}^{MN}$($\mathbf{C}_{ij} \in \mathbb{C}^{MN \times MN}$) to its trace-compressed version:
\begin{equation}
\mathbf{T}(\mathbf{C}) \triangleq \left[ \operatorname{tr}(\mathbf{C}_{ij} \mathbf{A}_{\mathbf{c}_d}) \right]_{i,j=1}^{MN},
\label{eq:compression_operator}
\end{equation}

For the simplified models, the LMMSE estimate of the direct channel coefficients follows a unified expression:
\begin{equation}
    \hat{\mathbf{g}}_{s} = \mathbf{C}_{\mathbf{g}\mathbf{y}_{s}} \mathbf{C}_{\mathbf{y}_{s}}^{-1}\mathbf{y}_{fm},
    \label{equ:LMNMSESimplifiedModelPartialKnown}
\end{equation}
where the subscript $s$ denotes the specific simplified model (i.e., $fr$, $iN$, or $dN$). While the general form remains consistent across models, their individual realizations differ in the covariance structures. Below, we provide the model-specific covariance matrices:
\begin{subequations}
\begin{align}
    \mathbf{C}_{\mathbf{g}\mathbf{y}_{fr}} =& \mathbf{C}_{\mathbf{g}\mathbf{y}_{iN}} = \mathbf{C}_{\mathbf{g}} \mathbf{B}_p^H, \\
    \mathbf{C}_{\mathbf{g}\mathbf{y}_{iN}} =& \mathbf{C}_{\mathbf{g}} \mathbf{B}_p^H + \mathbf{C}_{\tilde{\mathbf{h}}\mathbf{g}}^H \mathbf{A}_p^H;\\
    \mathbf{C}_{\mathbf{y}_{fr}} =& \mathbf{B}_p \mathbf{C}_{\mathbf{g}} \mathbf{B}_p^H
        + \sigma_d^2 \mathbf{A}_{\mathbf{c}_d} \odot \mathbf{C}_{\mathbf{g}} + \mathbf{C}_{\mathbf{n}}, \label{equ:gfr} \\
    \mathbf{C}_{\mathbf{y}_{iN}} =& \mathbf{B}_p \mathbf{C}_{\mathbf{g}} \mathbf{B_p}^H
        + \sigma_d^2 \mathbf{A}_{\mathbf{c}_d} \odot \mathbf{C}_{\mathbf{g}} + \mathbf{C}_{\boldsymbol{\delta}} + \mathbf{C}_{\mathbf{n}}  , \label{equ:giN} \\
    \mathbf{C}_{\mathbf{y}_{dN}} =& \mathbf{B}_p \mathbf{C}_{\mathbf{g}} \mathbf{B_p}^H
    + \sigma_d^2 \mathbf{A}_{\mathbf{c}_d} \odot \mathbf{C}_{\mathbf{g}} +    \nonumber \\
    &\mathbf{A}_p \mathbf{C}_{\tilde{\mathbf{h}}\mathbf{g}} \mathbf{B}_p^H + \mathbf{B}_p \mathbf{C}^H_{\tilde{\mathbf{h}}\mathbf{g}} \mathbf{A}_p^H +  \nonumber \\
    & \sigma_d^2 \mathbf{R} (\mathbf{C}_{\tilde{\mathbf{h}}\mathbf{g}}) + \sigma_d^2 \mathbf{R}(\mathbf{C}_{\tilde{\mathbf{h}}\mathbf{g}})^H +  \mathbf{C}_{\boldsymbol{\delta}} + \mathbf{C}_{\mathbf{n}}, \label{equ:gdN}
\end{align}
\end{subequations}
where $\mathbf{A}_{\mathbf{c}_d}=\mathrm{Diag}(\mathbf{c}_d)$, $\mathbf{C}_{\boldsymbol{\delta}}  = \mathbf{A}_p \mathbf{C}_{\tilde{\mathbf{h}}} \mathbf{A}_p^H + \sigma^2_{d} \mathbf{T}(\mathbf{C}_{\tilde{\mathbf{h}}})$ and $\mathbf{R}(\cdot)$ is a dimension-reduction operator that maps the input block vector matrix $\mathbf{C} \in \mathbb{C}^{(MN)^2 \times MN}$ to an $MN \times MN$ matrix by selective element extraction and weighting:
\begin{equation}
\mathbf{R}(\mathbf{C}) \triangleq \left[ \mathbf{c}_{ij}[j] \cdot c_d[j] \right]_{i,j=1}^{MN}.
\label{eq:dim_reduction}
\end{equation}
Here, $\mathbf{C} = [\mathbf{c}_{ij}]_{i,j=1}^{MN}$ is partitioned into $MN \times MN$ column vectors $\mathbf{c}_{ij} \in \mathbb{C}^{MN}$, notation $\mathbf{c}_{ij}[j]$ and $\mathbf{c}_d[j]$ means selecting the $j$-th element of vector $\mathbf{c}_{ij}$ and $\mathbf{c}_d$, respectively.

\subsection{LMMSE Sensing with Known Transmitted Symbols}
When the receiver has full knowledge of the transmitted symbols, the LMMSE estimates of the channel coefficients are given by:
\begin{subequations}
\begin{align}
\hat{\underline{\mathbf{h}}}_{fm} &  = \underline{\mathbf{C}}_{\mathbf{h}\mathbf{y}_{fm}} \underline{\mathbf{C}}_{\mathbf{y}_{fm}}^{-1} \mathbf{y}_{fm}, \\
\hat{\underline{\mathbf{g}}}_{s} &= \underline{\mathbf{C}}_{\mathbf{g}\mathbf{y}_{s}} \underline{\mathbf{C}}_{\mathbf{y}_{s}}^{-1} \mathbf{y}_{fm}.
\end{align}
\end{subequations}
where the underline denotes the case that the ISAC receiver knows the transmitted symbols. The corresponding covariance matrices for each model are specified as follows:
\begin{equation}
\begin{array}{ll@{\quad}l}
\underline{\mathbf{C}} _{\mathbf{h}\mathbf{y}_{fm}} = \mathbf{C}_{\mathbf{h}} \mathbf{A}^H, &
\underline{\mathbf{C}} _{\mathbf{y}_{fm}} = \mathbf{A} \mathbf{C}_{\mathbf{h}} \mathbf{A}^H + \mathbf{C}_{\mathbf{n}}; \\
\underline{\mathbf{C}} _{\mathbf{g}\mathbf{y}_{fr}} = \underline{\mathbf{C}} _{\mathbf{g}\mathbf{y}_{iN}} =  \mathbf{C}_{\mathbf{g}} \mathbf{B}^H, &
\underline{\mathbf{C}} _{\mathbf{g}\mathbf{y}_{dN}} = \mathbf{C}_{\mathbf{g}} \mathbf{B}^H + \mathbf{C}_{\tilde{\mathbf{h}}\mathbf{g}}^H \mathbf{A}^H, \\
\underline{\mathbf{C}} _{\mathbf{y}_{fr}} = \mathbf{B} \mathbf{C}_{\mathbf{g}} \mathbf{B}^H + \mathbf{C}_{\mathbf{n}}, &
\underline{\mathbf{C}} _{\mathbf{y}_{iN}} = \mathbf{B} \mathbf{C}_{\mathbf{g}} \mathbf{B}^H + \mathbf{C}_{\boldsymbol{\delta}} + \mathbf{C}_{\mathbf{n}},\\
\multicolumn{2}{l}{\underline{\mathbf{C}} _{\mathbf{y}_{dN}} = \mathbf{B} \mathbf{C}_{\mathbf{g}} \mathbf{B}^H + \mathbf{A} \mathbf{C}_{\tilde{\mathbf{h}}\mathbf{g}} \mathbf{B}^H + \mathbf{B} \mathbf{C}_{\mathbf{g}\tilde{\mathbf{h}}} \mathbf{A}^H + \mathbf{C}_{\boldsymbol{\delta}} + \mathbf{C}_{\mathbf{n}}.}
\end{array}
\nonumber
\end{equation}
% \begin{equation}
% \begin{aligned}
% \underline{\mathbf{C}} _{\mathbf{h}\mathbf{y}_{fm}} =& \mathbf{C}_{\mathbf{h}} \mathbf{A}^H, &
% \underline{\mathbf{C}} _{\mathbf{y}_{fm}} =& \mathbf{A} \mathbf{C}_{\mathbf{h}} \mathbf{A}^H + \mathbf{C}_{\mathbf{n}} \\
% \underline{\mathbf{C}} _{\mathbf{g}\mathbf{y}_{fr}} =& \underline{\mathbf{C}} _{\mathbf{g}\mathbf{y}_{iN}} =  \mathbf{C}_{\mathbf{g}} \mathbf{B}^H , 
% & \underline{\mathbf{C}} _{\mathbf{g}\mathbf{y}_{dN}} =& \mathbf{C}_{\mathbf{g}} \mathbf{B}^H + \mathbf{C}_{\tilde{\mathbf{h}}\mathbf{g}}^H \mathbf{A}^H\\
% \underline{\mathbf{C}} _{\mathbf{y}_{fr}} =& \mathbf{B} \mathbf{C}_{\mathbf{g}} \mathbf{B}^H + \mathbf{C}_{\mathbf{n}}, &
% \underline{\mathbf{C}} _{\mathbf{y}_{iN}} =& \mathbf{B} \mathbf{C}_{\mathbf{g}} \mathbf{B}^H
%     +  \mathbf{C}_{\boldsymbol{\delta}} + \mathbf{C}_{\mathbf{n}}, \\

% \underline{\mathbf{C}} _{\mathbf{y}_{dN}} = \mathbf{B} \mathbf{C}_{\mathbf{g}} \mathbf{B}^H + \mathbf{A} \mathbf{C}_{\tilde{\mathbf{h}}\mathbf{g}} \mathbf{B}^H + \mathbf{B} \mathbf{C}_{\mathbf{g}\tilde{\mathbf{h}}} \mathbf{A}^H + \mathbf{C}_{\boldsymbol{\delta}} + \mathbf{C}_{\mathbf{n}}. 
% \end{aligned}
% \nonumber
% \end{equation}

\subsection{LMMSE Sensing Error Analysis}
Compared to the LMMSE estimate of the full model, simplified models estimate only the direct channel coefficients. To evaluate the overall channel estimation error and its impact on channel capacity, the estimated channel from these three approaches must be zero-padded to construct a channel vector with the same dimensionality as that obtained from the full model. The zero-padding operation can be expressed as:
\begin{equation}
    \hat{\mathbf{h}}_{s}[i] = 
    \begin{cases} 
    \hat{\mathbf{g}}_{s}[k] & \text{if } i = k + MN(k-1), k =1,\dots, MN, \\ 
    0 & \text{otherwise}. 
    \end{cases}
    \label{equ:ZeroPadding}
\end{equation}
Using \eqref{equ:ZeroPadding}, the zero-padded channel estimates for the simplified models can be formed as $\hat{\mathbf{h}}_{fr}$, $\hat{\mathbf{h}}_{iN}$ and $\hat{\mathbf{h}}_{dN}$

The estimation error is defined as the deviation from the true channel coefficients: $\mathbf{e} = \mathbf{h} - \hat{\mathbf{h}}$. Given that $\mathbb{E}[\mathbf{h}]=0$ and $\mathbb{E}[\hat{\mathbf{h}}]=0$, the covariance matrix of the estimation error is
\begin{equation}
    \mathbf{C_{e}} = \mathbf{C_{h}} - \mathbf{C}_{\mathbf{h}\hat{\mathbf{h}}} - \mathbf{C}_{\hat{\mathbf{h}}\mathbf{h}} + \mathbf{C}_{\hat{\mathbf{h}}}.
    \label{equ:ErrorCovarianceMatrix}
\end{equation}
For the full model, substituting \eqref{equ:LMNMSEFullModelPartialKnown} into \eqref{equ:ErrorCovarianceMatrix} yields:
\begin{equation}
    \mathbf{C}_{\mathbf{e}_{fm}} = \mathbf{C}_{\mathbf{h}}-\mathbf{C}_{\mathbf{h}\mathbf{y}_{fm}} \mathbf{C}_{\mathbf{y}_{fm}}^{-1} \mathbf{C}_{\mathbf{h}\mathbf{y}_{fm}}^H.
\end{equation}
For the simplified models, the cross-covariance $\mathbf{C}_{\hat{\mathbf{h}}_{s}\mathbf{h}}$ and $\mathbf{C}_{\hat{\mathbf{h}}_{s}}$ are obtained by zero-padding $\mathbf{C}_{\hat{\mathbf{g}}_{s} \mathbf{h}}$ and $\mathbf{C}_{\hat{\mathbf{g}}_{s} }$. Their relationship can be expressed as
\begin{equation}
    \mathbf{C}_{\hat{\mathbf{h}}_{s} \mathbf{h}}[i,:] = 
    \begin{cases} 
    \mathbf{C}_{\hat{\mathbf{g}}_{s} \mathbf{h}}[k,:] & \text{if } i = k + MN(k-1),  \\ 
    \mathbf{0}^\top & \text{otherwise}. 
    \end{cases}
\end{equation}
\begin{equation}
    \mathbf{C}_{\hat{\mathbf{h}}_{s}}[i,j] = 
    \begin{cases} 
    \mathbf{C}_{\hat{\mathbf{g}}_{s}}[k,l] & \begin{aligned}
        &\text{if } i = k + MN(k-1)  \\ 
        &\text{and }j = l + MN(l-1),
    \end{aligned} \\ 
    0 & \text{otherwise}. 
    \end{cases}
\end{equation}
where $\mathbf{0}^\top$ is a zero row vector of matching dimension. The underlying matrices$\mathbf{C}_{\hat{\mathbf{g}}_{s} \mathbf{h}}$ and $\mathbf{C}_{\hat{\mathbf{g}}_{s} }$  are given as
\begin{subequations}
\begin{align}
\mathbf{C}_{\hat{\mathbf{g}}_{s} \mathbf{h}} & = \mathbb{E}[ \mathbf{C}_{\mathbf{g}\mathbf{y}_{s}}\mathbf{C}_{\mathbf{y}_{s}}^{-1}\mathbf{y}_{fm} \mathbf{h}^H ] \nonumber \\
&= \mathbf{C}_{\mathbf{g}\mathbf{y}_{s}}\mathbf{C}_{\mathbf{y}_{s}}^{-1} \mathbf{C}_{\mathbf{y}_{fm}\mathbf{h}}, \\
\mathbf{C}_{\hat{\mathbf{g}}_{s}} & = \mathbf{C}_{\mathbf{g}\mathbf{y}_{s}}\mathbf{C}_{\mathbf{y}_{s}}^{-1} \mathbf{C}_{\mathbf{y}_{fm}} \mathbf{C}_{\mathbf{y}_{s}}^{-1}\mathbf{C}_{\mathbf{g}\mathbf{y}_{s}}^H.
\end{align}
\end{subequations}
The error covariance matrix $\mathbf{C}_{\mathbf{e}_{s}}$ for each simplified model can then be computed using \eqref{equ:ErrorCovarianceMatrix}. Similarly, the error covariance matrices for the case that the ISAC receiver knows transmitted symbols, denoted as $\underline{\mathbf{C}}_{\mathbf{e}_{fm}}$ and $\underline{\mathbf{C}}_{\mathbf{e}_{s}}$, follow an analogous derivation.

We define the \textit{sensing error} as the total estimation error across all direct channel coefficients. For the case of partially known transmitted symbols, the sensing error $\mathcal{D}$ is the sum of selected diagonal elements of the covariance matrix $\mathbf{C_{e}}$, given by:
\begin{equation}
\mathcal{D} = \sum_{k=1}^{MN} \left[ \mathrm{diag}(\mathbf{C_e}) \right]_{k + MN(k-1)},
\end{equation}
where $\mathrm{diag}(\cdot)$ extracts the diagonal elements and the subscript $k + MN(k-1)$ selects specific elements spaced by $MN$ intervals. For the case of known transmitted symbols, the sensing error $\underline{\mathcal{D}}$ is defined as:
\begin{equation}
\underline{\mathcal{D}} = \mathbb{E}_{\mathbf{d}}\left[ \sum_{k=1}^{MN} \left[ \mathrm{diag}(\underline{\mathbf{C}}_{\mathbf{e}}) \right]_{k + MN(k-1)} \right],
\end{equation}
where the expectation is taken over the data symbols $\mathbf{d}$.

\section{Channel Capacity with Imperfect Channel State Information}
\label{sec:ChannelCapacity}
To facilitate the derivation of channel capacity, we first transform the estimated channel vectors and estimation errors into matrix form. Specifically, the estimated channel vector $\hat{\mathbf{h}}_{fm}$ and $\hat{\mathbf{h}}_{s}$, and their corresponding estimation errors, $\mathbf{e}_{fm}$ and $\mathbf{e}_{s}$, are reshaped into matrices $\hat{\mathbf{H}}_{fm}$, $\hat{\mathbf{H}}_{s}$, $\mathbf{E}_{fm}$, $\mathbf{E}_{s}$.

The general form of the received symbols can be expressed as: 
$$
\begin{aligned}
    \mathbf{y}  & =( \hat{\mathbf{H}} + \mathbf{E} ) (\mathbf{x}_p + \mathbf{x}_d ) +\mathbf{n}. 
\end{aligned}
$$

Since $\mathbf{x}_p$ is known at the receiver, we can formulate an equivalent channel model:
\begin{equation}
   \begin{aligned}
        \tilde{\mathbf{y}}& = \mathbf{y} - \hat{\mathbf{H}} \mathbf{x}_p\\
        & = \hat{\mathbf{H}} \mathbf{D} \mathbf{d} + \mathbf{A} \mathbf{e} +\mathbf{n},
    \end{aligned} 
    \label{equ:CapacityChannelModel}
\end{equation}
where $\mathbf{D}$ is a column selection matrix such that $\mathbf{D} \mathbf{d}=\mathbf{x}_d$ and $\mathbf{D}\mathbf{D}^H=\mathbf{A}_{\mathbf{c}_d}$.

The ergodic capacity of this channel model is obtained by taking the expectation of the mutual information over the distribution of the estimated channel $\hat{\mathbf{H}}$:
\begin{equation}
    \begin{aligned}
        \mathcal{C}& =\mathbb{E}_{\hat{\mathbf{H}}} \left[ I(\mathbf{d}; \tilde{\mathbf{y}} | \hat{\mathbf{H}}) \right].
    \end{aligned}
\end{equation}

\subsection{Capacity of full model} 
In the full model case, owing to the properties of LMMSE estimation, the estimation error $\mathbf{e}_{fm}$ is independent of $\hat{\mathbf{H}}_{fm}$. We denote the autocovariance matrix of $\mathbf{A} \mathbf{e}_{fm}$ as $\mathbf{K}_{\mathbf{e}_{fm}}$:
$$
\begin{aligned}
    \mathbf{K}_{\mathbf{e}_{fm}} &= \mathbb{E}[\mathbf{A} \mathbf{C}_{\mathbf{e}_{fm}} \mathbf{A}^H] =\mathbf{A}_p\mathbf{C}_{\mathbf{e}_{fm}}\mathbf{A}_p^H + \sigma_d^2 \mathbf{T}(\mathbf{C}_{\mathbf{e}_{fm}}).
\end{aligned}
$$

Therefore, the ergodic capacity of full model is
\begin{equation}
    \begin{aligned}
        \mathcal{C}_{fm}& = \mathbb{E}_{\hat{\mathbf{H}}_{fm}} \left[ \log_2 \det \left( \sigma_d^2 \hat{\mathbf{H}}_{fm} \mathbf{A}_{\mathbf{c}_d} \hat{\mathbf{H}}_{fm}^H \mathbf{K}_{\mathbf{z}_{fm}}^{-1} + \mathbf{I}_{MN} \right) \right],\\
    \end{aligned}
\end{equation}
where $\mathbf{K}_{\mathbf{z}_{fm}} = \mathbf{K}_{\mathbf{e}_{fm}} + \mathbf{C_n}$.

\paragraph{Upper Bound} Using the Jensen's inequality, an upper bound $\mathcal{C}^{\mathsf{U}}_{fm}$ on the ergodic capacity of full model can be found
$$
\begin{aligned}
    \mathcal{C}_{fm}& \leqslant  \log_2 \det \left( \sigma_d^2 \mathbb{E}_{\hat{\mathbf{H}}_{fm}}[ \hat{\mathbf{H}}_{fm} \mathbf{A}_{\mathbf{c}_d} \hat{\mathbf{H}}_{fm}^H ] \mathbf{K}_{\mathbf{z}_{fm}}^{-1} + \mathbf{I}_{MN} \right) \\
& = \log_2 \det \left( \sigma_d^2 \mathbf{T}(\mathbf{C}_{\hat{\mathbf{h}}_{fm}}) \mathbf{K}_{\mathbf{z}_{fm}}^{-1} + \mathbf{I}_{MN} \right),
\end{aligned}
$$
where $\mathbf{C}_{\hat{\mathbf{h}}_{fm}} = \mathbf{C}_{\mathbf{h}\mathbf{y}_{fm}} \mathbf{C}_{\mathbf{y}_{fm}}^{-1} \mathbf{C}_{\mathbf{h}\mathbf{y}_{fm}}^H$.
\begin{figure*}[htbp]
    \centering
    \includegraphics[width=0.95\linewidth]{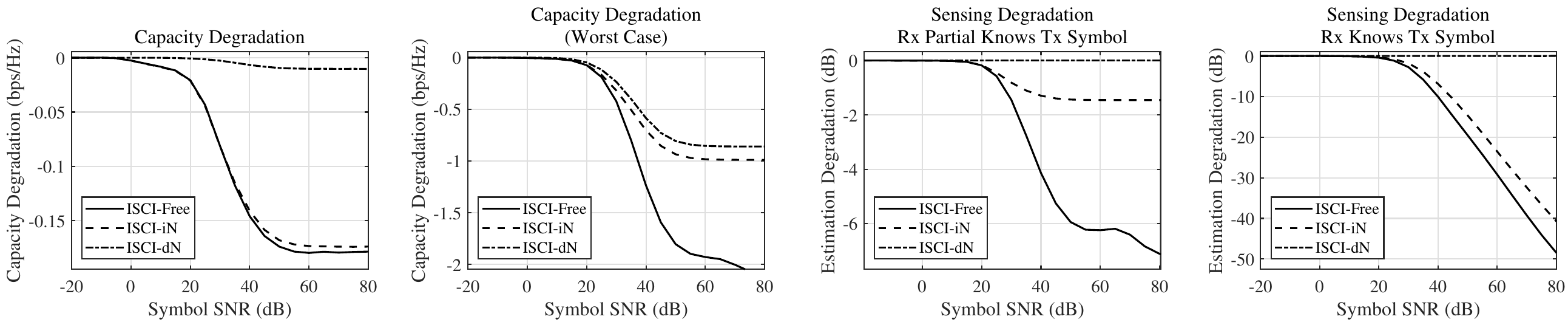}
    \caption{Sensing and communication performance degradation versus symbol SNR with fixed spread factor $0.001$.}
    \label{fig:FixSpreadDegradation}
\end{figure*}

\begin{figure*}[htbp]
\vspace{-4mm}
    \centering
    \includegraphics[width=0.95\linewidth]{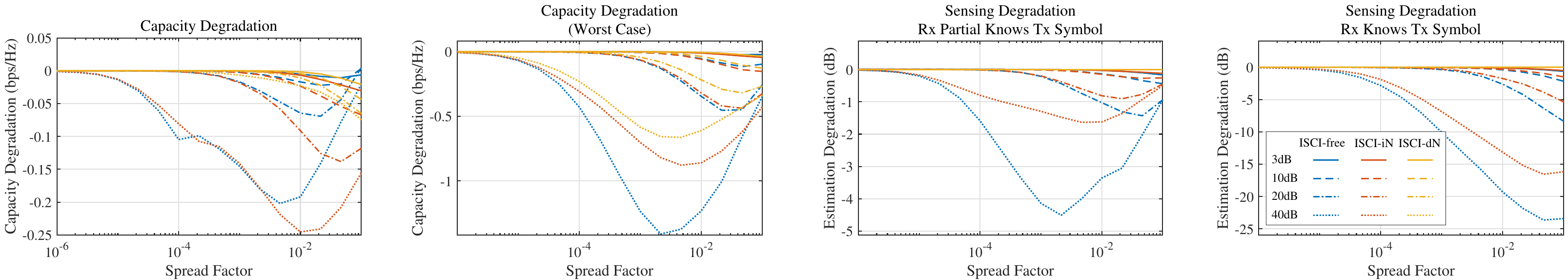}
    \caption{Sensing and communication performance degradation versus spread factor $\Delta_{\mathcal{D}}$ at fixed SNR values.}
    \label{fig:FixSNRDegradation}
\end{figure*}
\subsection{Capacity of simplified models} 
The expression for the ergodic capacity maintains the same general form across simplified models. However, since the LMMSE channel estimates in these models consider only the direct channel coefficients, the error term in \eqref{equ:CapacityChannelModel} is not independent of the channel estimate $\hat{\mathbf{H}}_{s}$(or equivalently, $\hat{\mathbf{g}}_{s}$). The conditional mutual information in the ergodic capacity can be rewritten as:
$$
\begin{aligned}
I(\mathbf{d}; \tilde{\mathbf{y}}_{s} | \hat{\mathbf{g}}_{s}) & = h(\mathbf{d}|\hat{\mathbf{g}}_{s}) - h(\mathbf{d}|\tilde{\mathbf{y}}_{s}, \hat{\mathbf{g}}_{s}).
\end{aligned}
$$

% Since $\hat{\mathbf{g}}_{s}$ is independent of $\mathbf{d}$, the differential entropy is given as \cite[p. 249, Th. 8.4.1]{coverElementsInformationTheory2005}:
Since $\hat{\mathbf{g}}_{s}$ is independent of $\mathbf{d}$, the differential entropy is given as
$$
h(\mathbf{d}|\hat{\mathbf{g}}_{s}) = \log_2\left( (2\pi e)^{L_d}\det (\sigma^2_d\mathbf{I}_{L_d}) \right).
$$

The differential entropy of $\mathbf{d}|\tilde{\mathbf{y}}_{s}, \hat{\mathbf{g}}_{s}$ can be obtained by calculating its conditional covariance matrix
\begin{equation}
\mathbf{C}_{\mathbf{d}|\tilde{\mathbf{y}}_{s}, \hat{\mathbf{g}}_{s}} = \sigma^2_d\mathbf{I}_{L_d} - \mathbf{C}_{\mathbf{d}\tilde{\mathbf{y}}_{s}|\hat{\mathbf{g}}_{s}} \mathbf{C}_{\tilde{\mathbf{y}}_{s}|\hat{\mathbf{g}}_{s}}^{-1} \mathbf{C}_{\mathbf{d}\tilde{\mathbf{y}}_{s}|\hat{\mathbf{g}}_{s}}^H,
\nonumber
\end{equation}
where
\begin{subequations}
   \begin{align}
    \mathbf{C}_{\mathbf{d}\tilde{\mathbf{y}}_{s}|\hat{\mathbf{g}}_{s}} &= \sigma^2_d\mathbf{I}_{L_d} \mathbf{D}^H \hat{\mathbf{H}}_{s}^H \nonumber ,\\
    \mathbf{C}_{\tilde{\mathbf{y}}|\hat{\mathbf{g}}_{s}} &= \sigma^2_d\hat{\mathbf{H}}_{s} \mathbf{A}_{\mathbf{c}_d} \hat{\mathbf{H}}_{s}^H + \mathbf{K}_{\mathbf{z}_{s}} \nonumber ,\\
    \mathbf{K}_{\mathbf{z}_{s}} &= \mathbb{E} \left[ \mathbf{A} \mathbf{C}_{\mathbf{e}_{s}|\hat{\mathbf{g}}_{s}} \mathbf{A}^H \right] + \mathbf{C_{n}} \nonumber ,\\
    &= \mathbf{A}_p \mathbf{C}_{\mathbf{e}_{s}|\hat{\mathbf{g}}_{s}} \mathbf{A}^H_p + \sigma_d^2 \mathbf{T}(\mathbf{C}_{\mathbf{e}_{s}|\hat{\mathbf{g}}_{s}}) + \mathbf{C_{n}}. \nonumber 
\end{align} 
\end{subequations}
Here, $\mathbf{C}_{\mathbf{e}_{s}|\hat{\mathbf{g}}_{s}}$ is the conditional autocovariance matrix of estimation error, representing the residual uncertainty after subtracting the components estimable from $\hat{\mathbf{g}}_{s}$ via LMMSE and it is given by
\begin{equation}
\begin{aligned}
    \mathbf{C}_{\mathbf{e}_{s}|\hat{\mathbf{g}}_{s}} &= \mathbf{C}_{\mathbf{e}_{s}} - \mathbf{C}_{\mathbf{e}_{s}\hat{\mathbf{g}}_{s}}\mathbf{C}^{-1}_{\hat{\mathbf{g}}_{s}}\mathbf{C}_{\hat{\mathbf{g}}_{s}\mathbf{e}_{s}} \\
    &= \mathbf{C}_{\mathbf{e}_{s}} -  (\mathbf{C}_{\mathbf{h}\hat{\mathbf{g}}_{s}} -\mathbf{C}_{\hat{\mathbf{g}}_{s}}) \mathbf{C}^{-1}_{\hat{\mathbf{g}}_{s}} ( \mathbf{C}_{\hat{\mathbf{g}}_{s}\mathbf{h}} - \mathbf{C}_{\hat{\mathbf{g}}_{s}}).
\end{aligned}
\nonumber
\label{equ:ConditionalCovariance}
\end{equation}

Thus, the ergodic capacity for the simplified models is:
\begin{equation}
    \begin{aligned}
        & \mathcal{C}_{s} =\mathbb{E}_{\hat{\mathbf{H}}_{s}} \left[ \log_2 \det \left(\sigma^2_d\mathbf{I}_{L_d} \mathbf{C}_{\mathbf{d}|\tilde{\mathbf{y}}_{s}, \hat{\mathbf{g}}_{s}}^{-1} \right) \right] \\
        & \overset{(a)}{=} \mathbb{E}_{\hat{\mathbf{H}}_{s}} \left[ \log_2 \det \left( \sigma^2_d\hat{\mathbf{H}}_{s} \mathbf{A}_{\mathbf{c}_d} \hat{\mathbf{H}}_{s}^H \mathbf{K}_{\mathbf{z}_{s}}^{-1} + \mathbf{I}_{MN} \right) \right],
    \end{aligned}
    \label{equ:ErgodicCapacitySimplifiedModel}
\end{equation}
where step (a) is obtained through Woodbury matrix identity and Sylvester's determinant identity.

\paragraph{Upper Bound} Using the Jensen's inequality, an upper bound $\mathcal{C}^{\mathsf{U}}_{s}$ is obtained:
$$
\begin{aligned}
\mathcal{C}_{s} & \leqslant \log_2 \det \left( \sigma_{d}^2 \mathbb{E}_{\hat{\mathbf{H}}_{s}}[\hat{\mathbf{H}}_{s} \mathbf{A}_{\mathbf{c}_d} \hat{\mathbf{H}}_{s}^H] \mathbf{K}_{\mathbf{z}_{s}}^{-1} + \mathbf{I}_{MN} \right) \\
    & = \log_2 \det \left( \sigma_{d}^2 \mathrm{Diag}(\mathbf{b}_{s}) \mathbf{K}_{\mathbf{z}_{s}}^{-1} + \mathbf{I}_{MN} \right)
\end{aligned}
$$
where $\mathbf{b}_{s}=[\mathbf{c}_d[1] \sigma^2_{\hat{\mathbf{g}}}[1], \mathbf{c}_d[2] \sigma^2_{\hat{\mathbf{g}}}[2], \dots, \mathbf{c}_d[MN] \sigma^2_{\hat{\mathbf{g}}}[MN]]^T$ and $\sigma^2_{\hat{\mathbf{g}}}[i]$ is the $i$-th diagonal element of $\mathbf{C}_{\hat{\mathbf{g}}_{s}}$.

\paragraph{Lower Bound (Worst Case)} The capacity in \eqref{equ:ErgodicCapacitySimplifiedModel} corresponds to the best-case scenario where the LMMSE estimates $\hat{\mathbf{g}}_{s}$ are used to cancel the predictable portion of the channel estimation error $\mathbf{A}\mathbf{e}_{s}$. In many practical systems, however, this cancellation is not performed. If we assume that $\mathbf{A}\mathbf{e}_{s}$ is independent from $\hat{\mathbf{g}}_{s}$, a lower bound $\mathcal{C}^{\mathsf{L}}_{s}$ representing the worst-case scenario can be derived.

Mathematically, from \eqref{equ:ConditionalCovariance}, we have $\mathbf{C}_{\mathbf{e}_{s}} \succeq \mathbf{C}_{\mathbf{e}_{s}|\hat{\mathbf{g}}_{s}}$. Let $\check{\mathbf{K}}_{\mathbf{z}_{s}} = \mathbb{E} \left[ \mathbf{A} \mathbf{C}_{\mathbf{e}_{s}} \mathbf{A}^H \right] + \mathbf{C_{n}} $; it follows that $\check{\mathbf{K}}_{\mathbf{z}_{s}}^{-1} \preceq  \mathbf{K}_{\mathbf{z}_{s}}^{-1}$. The lower bound is then:
$$
\mathcal{C}_{s} \geqslant \mathbb{E}_{\hat{\mathbf{H}}_{s}} \left[ \log_2 \det \left( \sigma^2_d\hat{\mathbf{H}}_{s} \mathbf{A}_{\mathbf{c}_d} \hat{\mathbf{H}}_{s}^H \check{\mathbf{K}}_{\mathbf{z}_{s}}^{-1} + \mathbf{I}_{MN} \right) \right],
$$
where $\check{\mathbf{K}}_{\mathbf{z}_{s}} = \mathbf{A}_p \mathbf{C}_{\mathbf{e}_{s}} \mathbf{A}^H_p + \sigma_d^2 \mathbf{T}(\mathbf{C}_{\mathbf{e}_{s}}) + \mathbf{C_{n}}$.

\section{Performance Degradation of Simplified Model and Numerical Experiment}
In this section, we present numerical simulations to validate the theoretical results for both communication and sensing performance derived in Sections~\ref{sec:LMMSESensing} and~\ref{sec:ChannelCapacity}. Additionally, we quantitatively evaluate the performance degradation of the simplified models relative to the full model in terms of sensing error and channel capacity.

We consider an ISAC system operating over a time-frequency grid of size $M=6$ and $N=4$. An OFDM waveform without guard intervals is adopted, corresponding to $TF=1$, $F=30\text{kHz}$, with both transmit and receive prototype pulses being rectangular windows of length $T$. The pilot symbols' power is set equal to that of data symbols, i.e., $\sigma_{p}^2=\sigma^2_d$. Pilots and data are equally spaced on the time-frequency plane with a pilot density of 25\%. The channel scattering function is assumed to be constant over its support, denoted by $\sigma_{\mathcal{D}}$.

\begin{figure}[htbp]
    \centering
    \includegraphics[width=1\linewidth]{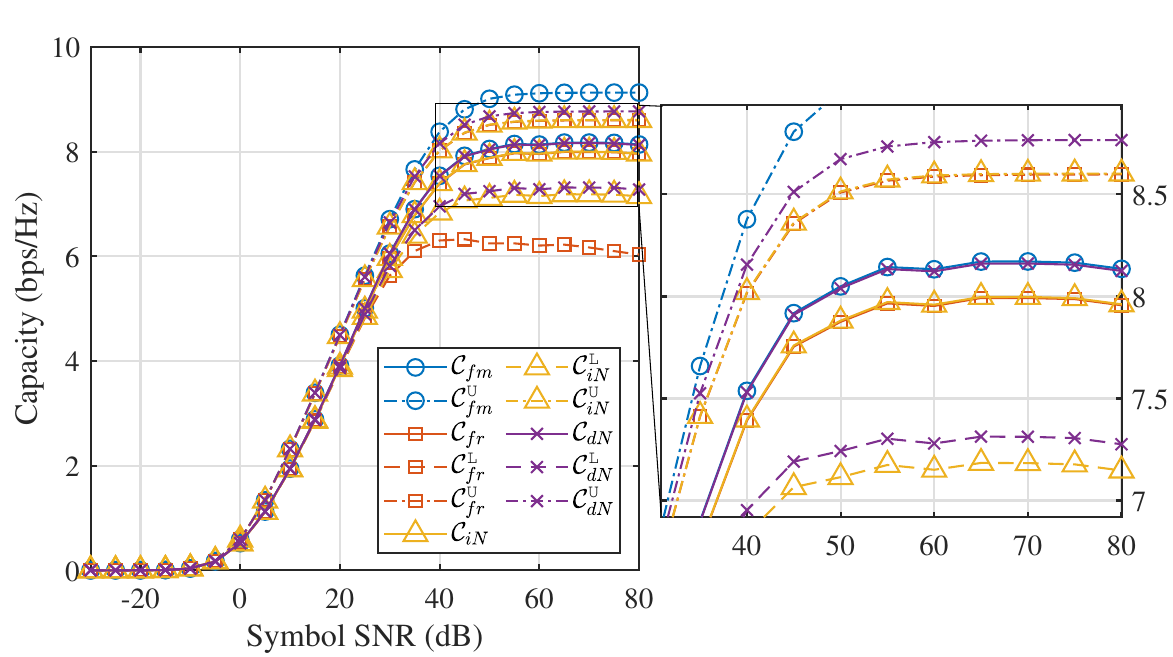}
    \caption{Ergodic capacity versus symbol SNR.}
    \label{fig:FixSpreadCapacity}
\end{figure}

\begin{figure}
    \flushleft
    \includegraphics[width=1\linewidth]{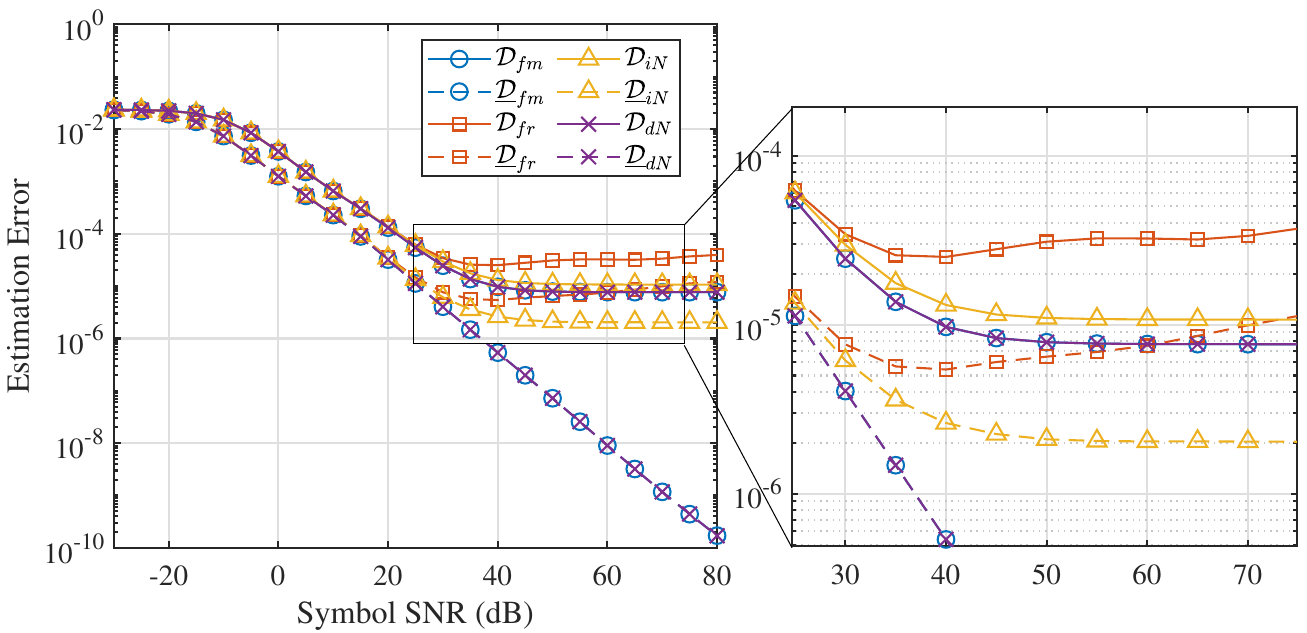}
    \caption{Sensing error versus symbol SNR.}
    \label{fig:FixSpreadSensingError}
\end{figure}

Fig.~\ref{fig:FixSpreadCapacity} and Fig.~\ref{fig:FixSpreadSensingError} show the numerical results of the ergodic capacity (with corresponding bounds) and sensing error, respectively, obtained with a fixed spread factor $\Delta_{\mathcal{D}}=0.001$ and varying symbol SNR\footnote{Symbol SNR is defined as $\frac{\mathrm{tr}(\mathbf{C_{h}}) \sigma^2_d}{MN \sigma^2_n}$}. In high SNR regimes, the performance of pilot-assisted sensing and communication reaches a plateau because the use of spaced pilots limits further channel estimation improvement. For the Tx symbols known case, the estimation error of full model and ISCI-dN model do not experience a plateau.

Fig.~\ref{fig:FixSpreadDegradation} illustrates the performance degradation of simplified models relative to the full model, showing: (a) capacity difference $\mathcal{C}_{s}-\mathcal{C}_{fm}$, (b) worst case capacity difference $\mathcal{C}^{\mathsf{L}}_s-\mathcal{C}_{fm}$, (c) normalized pilot-aided sensing error $\mathcal{D}_{s}/\mathcal{D}_{fm}$, and (d) normalized sensing error with known symbols $\underline{\mathcal{D}}_{s}/\underline{\mathcal{D}}_{fm}$. The sensing performance of the ISCI-dN model is nearly identical to that of the full model. From low to moderate SNR regimes, the performance differences among the four models are negligible.

Fig.~\ref{fig:FixSNRDegradation} demonstrates the performance degradation of simplified models with varying spread factor at four fixed symbol SNR values (3dB, 10dB, 20dB, and 40dB). Generally, as the spread factor increases, the performance loss in Fig.~\ref{fig:FixSNRDegradation}(a), Fig.~\ref{fig:FixSNRDegradation} (b), and Fig.~\ref{fig:FixSNRDegradation} (c) initially increases and then decreases. This trend suggests that at both lower and higher spread factors, the performance loss may remain within an acceptable range. However, Fig.~\ref{fig:FixSpreadDegradation} (d) and Fig.~\ref{fig:FixSNRDegradation} (d) reveal that regardless of the channel spread, the performance degradation of simplified models at high SNR is significant and must be considered in practical system design.

\section{Conclusion}

In this paper, we presented a systematic framework for characterizing the impact of ISCI on the performance of ISAC systems operating over doubly dispersive channels. Our numerical results demonstrate a fundamental trade-off between modeling accuracy and computational complexity. Key findings include:
\begin{itemize}
    \item The ISCI-dN model closely approximates the full model in sensing accuracy, making it a promising candidate for practical implementations.
    \item Performance loss is most pronounced in high-SNR scenarios, where simplified models fail to compensate ISCI.
    \item At low to moderate SNR or in mildly dispersive channels, all simplified models perform acceptably, justifying their use for complexity reduction.
\end{itemize}

\bibliographystyle{IEEEtran}
\bibliography{ref}

\end{document}